\documentstyle[12pt]{article}
\def\be{\begin{equation}}
\def\ee{\end{equation}}
\def\bea{\begin{eqnarray}}
\def\eea{\end{eqnarray}}

\begin{document}

\begin{center}
{\Large{\bf Moving Branes in Presence of the Background Tachyon
Fields }}

\vskip .5cm {Zahra Rezaei and Davoud Kamani}
\vskip .1cm
{\it Physics Department, Amirkabir University of Technology
(Tehran Polytechnic)\\
P.O.Box: 15875-4413, Tehran, Iran}\\
{\sl e-mails: z.rezaei , kamani@aut.ac.ir}\\
\end{center}

\begin{abstract}

We compute the boundary state associated with a moving D$p$-brane
in the presence of the open string tachyon field as a background
field. The effect of the tachyon condensation on the boundary
state will be discussed. It leads to a boundary state associated
with a lower dimensional moving D-brane or a stationary
instantonic D-brane. The former originates from condensation
along the spatial directions and the latter comes from the temporal
direction of the D-brane's worldvolume.
Using the boundary state we also study the interaction amplitude
between two arbitrary D$p_1$ and D$p_2$-branes. The long range
behavior of the amplitude is investigated which shows an obvious
deviation from the conventional form, due to the presence of the
background tachyon field.

\end{abstract}



\vskip .5cm
\newpage

\section{Introduction}
Open string tachyon can be considered as instabilities of the
branes, because open strings introduce the quantum excitations of
the branes \cite{1}. Tachyon potential has a stationary point where
the negative potential energy of the tachyon cancels the tension of
the D-brane \cite{2}. This process which is called tachyon
condensation ends when the brane has completely disappeared. During
the condensation process lower dimensional branes are produced
\cite{3,4}.

On the other hand we have the boundary state as a quantum state,
which contains closed string states \cite{5}. It can be used to
study D-branes. In the other words, a D-brane couples to all
states of the closed string via the boundary state. Thus, we can
suppose that the exchange of closed strings between two D-branes
is responsible for the branes interaction. For calculating it we
can just connect their corresponding boundary states through the
closed string propagator. The coherent state method \cite{6} and
the path integral approach \cite{7, 8} have been used to obtain
boundary state. Furthermore, the boundary state in the presence of
background fields such as $B_{\mu\nu}$ and $U(1)$ gauge fields in
the compact spacetime \cite{9}, and in the presence of the tachyon
field \cite{10, 11} have been investigated.

Apart from the background $U(1)$ gauge field and the open string
tachyon field, which are parallel to the brane's worldvolume, the
transverse fluctuations of a D-brane also are essential to study
it as a dynamical object. Scarcity of this kind of multilateral
discussion motivated us to follow this process in this paper in
spite of some technical difficulties. Beside, usually in the
literature a full brane in the presence of a one-dimensional
background tachyon field has been considered and the effect of
one-stage condensation on that brane has been studied. But here we
study a D-brane with an arbitrary dimension. Therefore, the
Dirichlet boundary conditions also will be present. In our set-up
the tachyon field has components along all the directions of the
D-brane's worldvolume. This tachyon profile leads to various
condensations and hence variety of the resulted branes.

In the present article by using the path integral approach we
calculate the boundary state corresponding to a moving D$p$-brane
in the presence of a tachyon field. Consequently, we obtain the
disk partition function of the closed string. The effect of the
tachyon condensation on this partition function will be studied.
The condensation is applied along the spatial worldvolume
directions and gives a partition function associated with a moving
lower dimensional D-brane. Difference with the conventional
tachyon condensation (e.g. see \cite{10}) is the presence of a
tachyon dependent factor in the resulted partition function. That
is, although the brane's dimension decreases, the effect of the
tachyon does not remove by condensation. In this process the
transverse fluctuations of a D$p$-brane prevent the normal tachyon
condensation from occurring. Applying the condensation along the
temporal direction of the D$p$-brane's worldvolume gives an
instantonic stationary brane.

The final goal of the paper is applying the boundary state to
obtain interaction amplitude between two moving D-branes and
study its behavior for large distances of the branes. We observe
that due to the inclusion of the open string tachyon background
(which is equivalent to consider instability of the bosonic
D-branes), the long range interaction of the branes goes to zero.
This is a consequence of rolling of the tachyon toward its
minimum potential. We observe that for the interaction of two
D-instantons the conventional long time interaction amplitude
restores.
%%%%%%%%%%%%%%%%%%%%%%%%%%%%%%%%%%%%%%%%%%%%%%%%%%%%%%%%%%%%%%
\section{Boundary state and tachyon condensation}

To determine the boundary state, associated with a moving D$p$-brane
in the presence of the tachyon field, we begin with an appropriate
sigma-model for the string. This action contains the bulk
term, a tachyonic term on the boundary and a velocity term
corresponding to the motion of the brane
\begin{equation}
S_{\rm bulk}
=-\frac{1}{4\pi\alpha'} {\int}_{\Sigma} d^{2}\sigma
(\sqrt{-h}h^{ab}g_{\mu\nu}\partial_{a}X^{\mu}\partial_{b}X^{\nu}),
\end{equation}
\begin{equation}
S_{\rm boundary}={\int}_{\partial\Sigma} d\sigma
(V^{i}X^{0}\partial_{\tau}X^{i}+i
U_{\alpha\beta}X^{\alpha}X^{\beta}),
\end{equation}
where $\Sigma$ is the worldsheet of the closed string, exchanged
between the branes, $\partial\Sigma$ indicates the boundary of
this worldsheet, which can be at $\tau=0$ or $\tau=\tau_{0}$ and
the $d$-dimensional spacetime metric is
$g_{\mu\nu}=(-1,1,\cdot\cdot\cdot,1)$. In addition, we defined
$V^{i}=\frac{v^{i}}{2\pi\alpha'}$ where $v^{i}$ is the brane's
velocity component along $X^{i}$ direction. The coupling of the
tachyon field to the string via integration over the worldsheet
boundary has been discussed in \cite{12}. Since it occurs as
square in the action, i.e. $T^2(X)$, thus in order to produce a
Gaussian integral the tachyon profile must be chosen to be linear,
$T(X)=a+u_{\mu}X^{\mu}$. The constant $a$ has been shifted away in
(2). Note that we also consider the symmetric matrix $U$ to have
nonzero elements only along the worldvolume of the D$p$-brane. The
set $\{X^{\alpha}\}$ specifies the directions along the D$p$-brane
worldvolume and $\{X^{i}\}$ shows the directions perpendicular to
it.
%%%%%%%%%%%%%%%%%%%%%%%%%%%%%%%%%%%%%%%%%%%%%%%%%%%%%%%%%%%%%%
\subsection{The boundary state}

Now consider the mode expansion of the coordinates of the
closed string
\begin{equation}
X^{\mu}(\sigma,\tau)=x_{0}^{\mu}+2\alpha'p^\mu\tau
+\sqrt{\frac{\alpha'}{2}} \sum_{m>0}m^{-1/2}
(x^{\mu}_{m}e^{2im\sigma}+\overline{x}^{\mu}_{m} e^{-2im\sigma}),
\end{equation}
where we define $x$ and $\overline{x}$ as combinations of the
bosonic modes
\begin{eqnarray}
&~& x_{m}=a_{m}e^{-2im\tau}+\widetilde{a}_{m}^{\dagger}e^{2im\tau},
\nonumber\\
&~& \overline{x}_{m}=a_{m}^{\dagger}e^{2im\tau}
+\widetilde{a}_{m}e^{-2im\tau},
\end{eqnarray}
in which $a_{m}^{\mu}=\frac{i}{\sqrt{m}}\alpha_{m}^{\mu}$ and
$a_{m}^{\dagger\mu}=\frac{-i}{\sqrt{m}}\alpha_{-m}^{\mu}$. Similar
relations also hold for $\widetilde{a}_{m}^{\mu}$ and
$\widetilde{a}_{m}^{\dagger\mu}$. If we interpret the equations (4)
as eigenvalue equations \cite{7}, the corresponding eigenstate is
\begin{equation}
|x,\overline{x}\rangle=\prod_{m=1}^\infty\exp
\bigg{(}{-\frac{1}{2}\overline{x}_{m}x_{m}-
a_{m}^{\dagger}\widetilde{a}_{m}^{\dagger}+a_{m}^{\dagger}x_{m}+
\overline{x}_{m}\widetilde{a}_{m}^{\dagger}}\bigg{)}|vac\rangle,
\end{equation}
where contraction with the metric $g_{\mu\nu}$ is applied implicitly.
This state is the boundary state of the closed string due
to the bulk term of the string sigma-model without any boundary
interaction. Naturally deforming the action by adding non-vanishing
boundary contributions, leads to the deformed boundary state
\begin{equation}
|B;S_{\rm
boundary}\rangle=\int[dxd\overline{x}]e^{iS_{\rm
boundary}[x,\overline{x}]}|x,\overline{x}\rangle.
\end{equation}

The boundary actions related to the tachyon, $S_{T}$, and the
velocity term, $S_{V}$, can be written in terms of modes
\begin{equation}
S_{T}=i\pi
x_{0}^{\alpha}U_{\alpha\beta}x_{0}^{\beta}+i\pi\alpha'\sum_{m=1}^{\infty}
\overline{x}_{m}^{\alpha}\frac{U_{\alpha\beta}}{m}x_{m}^{\beta},
\end{equation}
\begin{equation}
S_{V}=v^{i}x_{0}^{0}p^{i}+iv^{i}
\sum_{m=1}^{\infty}\bigg{(}\frac{1}{2}(\overline{x}_{m}^{0}x_{m}^{i}+
\overline{x}_{m}^{i}x_{m}^{0})-\overline{x}_{m}^{0}a_{m}^{i}
-\widetilde{a}_{m}^{i}x_{m}^{0}\bigg{)}.
\end{equation}
From now on we impose a selected direction $X^{i_{0}}$ for the
motion of the D$p$-brane and define $v^{i_{0}}=v$. Substituting (7)
and (8) into (6), and also considering the contribution of the bulk
action in the boundary, gives the boundary state. The oscillating
part of this state is
\begin{equation}
|B_{x}\rangle^{\rm osc}=\prod_{m=1}^{\infty}\frac{1}{\det R_{(m)}}
\exp \bigg{(}\sum^\infty_{m=1} a_{m}^{\dagger}\cdot{\cal{S}}_{(m)}
\cdot\widetilde{a}_{m}^{\dagger}\bigg{)}|0\rangle,
\end{equation}
where
\begin{eqnarray}
\left\{
\begin{array}{rcl} &
R_{(m)ab}=-2\Omega_{ab}+\frac{2\pi\alpha'}{m}
U_{\alpha\beta}\delta_{\;\;a}^{\alpha}
\delta^{\beta}_{\;\;b},\\
& \Omega_{ab}=-\frac{1}{2}g_{ab}-\frac{1}{2}v(\delta_{\;\;a}^{0}
\delta^{i_{0}}_{\;\;b}+\delta_{\;\;a}^{i_{0}}\delta^{0}_{\;\;b}),
\end{array}\right.
\end{eqnarray}
and
\begin{equation}
{\cal{S}}_{(m)\mu\nu}=2(R_{(m)}^{-1})_{ab}
\delta_{\;\;\mu}^{a}\delta^{b}_{\;\;\nu}-g_{\mu\nu}.
\end{equation}
The indices $a$ and $b$ indicate worldvolume and motion directions
(i.e. $a, b\in\{\alpha, i_{0}\}$). It is seen that when the
velocity $v$ and the tachyon matrix $U$ are zero, there is
$(R_{(m)}^{-1})_{ab}=g_{ab}$. Hence the boundary state (9) reduces
to the state for a D$p$-brane where $X^{\alpha}$'s,
$\alpha=0,...,p$, and $X^{i}$'s, $i=p+1,...,d-1$, obey the Neumann
and Dirichlet boundary conditions, respectively \cite{13}.

The infinite product in (9) is generated by the path integral.
Zeta function regularization can be used to avoid this divergent
quantity \cite{14},
\begin{equation}
\prod_{m=1}^{\infty}\bigg{[}\det \bigg{(}-2\Omega+
2\pi\alpha'\frac{W}{m}\bigg{)}\bigg{]}^{-1}=
\sqrt{\det(-2\Omega)}\;\det\Gamma \bigg{(}
1-\frac{\pi\alpha'W}{\Omega}\bigg{)},
\end{equation}
where the matrix $W$ is defined by
$W_{ab}=U_{\alpha\beta}\delta_{\;\;a}^{\alpha}\delta^{\beta}_{\;\;b}$.

The zero mode part of the boundary state becomes
\begin{eqnarray}
|B_x \rangle ^0
&~& =\frac{T_p}{2\sqrt{\det U}}\;\int
dp^{\alpha}\;\exp(-\frac{1}{4\pi}
P^T U^{-1}P)\;\delta(x_{0}^{i_{0}}-vx_{0}^{0}-y^{i_{0}})
\prod_{i'\neq
i_{0}}\delta(x_{0}^{i'}-y^{i'})
\nonumber\\
&~& \times \prod_{\alpha}|p_{L}^{\alpha}=p_{R}^{\alpha}\rangle
\prod_{i'\neq i_{0}}|p_{L}^{i'}=p_{R}^{i'}=0
\rangle|p_{L}^{i_{0}}=p_{R}^{i_{0}}=\frac{1}{2}vp^{0}\rangle,
\end{eqnarray}
where the vector $P$ is defined by
$P_{\alpha}=vp^{i_{0}}\delta^{0}_{\;\alpha}-\frac{1}{2}p_{\alpha}$.
The momentum dependent exponential term appears due to the presence
of the momentum components in the zero mode parts of the boundary
actions. Two delta functions indicate the position of the brane.
After performing the integration over momenta
the matter part of the boundary state takes the form
\begin{eqnarray}
|B_{x}\rangle
&~&=|B_{x}\rangle^{\rm osc}|B_{x}\rangle^{0}
\nonumber\\
&~&=T_{p}\;\frac{\pi(4\pi)^p}{v^{2}+1/2}\;
\prod_{m=1}^{\infty}\frac{1}{\det R_{(m)}}\;\exp
\bigg{(}{\sum_{m=1}^{\infty}a_{m}^{\dagger}\cdot{\cal{S}}(m)\cdot
\widetilde{a}_{m}^{\dagger}}\bigg{)}\nonumber\\
&~&\times\delta(x_{0}^{i_{0}}-vx_{0}^{0}-y^{i_{0}}) \prod_{i'\neq
i_{0}}\delta(x_{0}^{i'}-y^{i'})|vac\rangle,
\end{eqnarray}
where $|vac\rangle=|0\rangle_{\alpha}|0\rangle_{\widetilde{\alpha}}|p\rangle$
is written in this form for briefness.
%%%%%%%%%%%%%%%%%%%%%%%%%%%%%%%%%%%%%%%%%%%%%%%%%%%%%%%%%%%%%%
\subsection{Partition function and tachyon condensation}

Since partition function is defined by ${\cal{Z}}=\int DX
e^{iS[X]}$, it is obvious that there exists a very natural
connection between boundary state and the partition function: the
latter is just given by the vacuum amplitude of the boundary state
\begin{equation}
{\cal{Z}}=\langle vac|B;S_{\rm boundary}\rangle.
\end{equation}
Therefore, the normalization factors in the Eq. (14) comes from
the disk partition function \cite{15} which can also be derived by
evaluating the string path integral on a disk
\begin{equation}
{\cal{Z}}_{\rm
disk}=T_{p}\;\frac{\pi(4\pi)^p}{v^{2}+1/2}\;
\prod_{m=1}^{\infty}\bigg{[}\det\bigg{(}g_{ab}
-v(\delta^{0}_{\;\;a}\delta^{i_{0}}_{\;\;b}
+\delta^{0}_{\;\;b}\delta^{i_{0}}_{\;\;a})
+\frac{2\pi\alpha'}{m}U_{\alpha\beta}
\delta^{\alpha}_{\;\;a}\delta^{\beta}_{\;\;b}\bigg{)}\bigg{]}^{-1}.
\end{equation}
Note that the disk diagram in the closed string
theory shows a propagating closed string from the boundary of the
disk, which then disappears.

Presence of the open string tachyon field as a background field in
our case, enables us to study the effect of tachyon condensation
on the partition function. In the case at hand where the tachyon
profile is linear, studying tachyon condensation equals to sending
the elements of the tachyon matrix $U$ to infinity \cite{10}.

Here our tachyon matrix possesses all elements along the brane
worldvolume. Remember that, $U_{\alpha\beta}$ is a $(p+1)\times(p+1)$
matrix. Without loss of generality let it be a
diagonal matrix. We consider condensation of all spatial components of $U$
which can be done one by one for each component or at once for all
of them. After successive condensations along the
spatial directions of the D$p$-brane
$\{X^{\overline{\alpha}}|{\bar \alpha=1,2, \cdot\cdot\cdot,p}\}$,
which the limit $U_{\overline{\alpha}\overline
{\alpha}}\rightarrow\infty$ is applied,
the partition function (16) becomes
\begin{equation}
{\cal{Z}}_{\rm
disk}=T_{p}\;\frac{\pi(4\pi)^p}{v^{2}+1/2}\;
(2\pi\sqrt{\alpha'})^{p}\sqrt{\det U'}
\prod_{m=1}^{\infty}\bigg{(}1-v^{2}+\frac{2\pi\alpha'}{m}U_{00}
\bigg{)}^{-1},
\end{equation}
where $U'$ is a new diagonal $p\times p$ tachyon matrix which does
not contain the element $U_{00}$. The zeta function
regularization, $\prod_{m=1}[\det(\frac{2\pi
\alpha'U'}{m})]^{-1}=(2\pi\sqrt{\alpha'})^{p}\sqrt{\det U'}$, has
been used in (17). The relation between D-branes' tensions,
$T_{p-q}=T_{p}\;(2\pi \sqrt{\alpha'})^{q}$, enables us to
interpret (17) as the partition function related to a moving
D$0$-brane with effective tension,
${\cal{T}}_{0}=T_{0}\;\frac{\pi(4\pi)^p} {v^{2}+1/2}\;\sqrt{\det
U'}$. This considerable difference with conventional tachyon
condensation \cite{10}, comes from the momentum dependent
exponential factor which exists due to the presence of zero modes
in both tachyon and velocity boundary actions. In the absence of
the velocity term there is no momentum dependence in partition
function and the factor $\frac{1}{\sqrt{\det U}}$ which appears
from zero modes in the tachyon action cancels out the factor
$\sqrt{\det U}$ which comes from the tachyon condensation in the
infinite determinant. However, an additional factor of $\sqrt{\det
U}$ appears because of Gaussian integration over momenta and leads
to this unusual behavior of partition function after tachyon
condensation.

As the next step, performing tachyon condensation along the
$X^{0}$-direction in Eq. (17), eliminates the velocity and results in a
D-instanton with the partition function of
\[{\cal{Z}}_{\rm
disk}=T_{(-1)}\;\frac{\pi(4\pi)^p} {v^{2}+1/2}\;\sqrt{\det U}.\]
In other words, temporal tachyon condensation fixes the D-brane in
time as well as eliminates its velocity and fixes it in the space.
Generally, temporal condensation on a moving D$p$-brane leads to a
stationary instantonic D$p$-brane (i.e. eliminates the time
direction of the worldvolume), and condensation of the spatial
components of the tachyon field also reduces the D$p$-brane
dimension.

Accordingly after tachyon condensation along any spatial direction
of the moving D$p$-brane's worldvolume, its dimension decreases by
one in such a manner that after $q$ successive condensations we
have a D$(p-q)$-brane in the presence of a
$U_{(p-q+1)\times(p-q+1)}$ tachyon field. The main difference with
the usual case is that although the brane's dimension decreases
but the effect of the tachyon remains in the root factor.

In the following section, by making use of the boundary state
formalism, we compute the interaction amplitude between two
D-branes in the closed string channel.
%%%%%%%%%%%%%%%%%%%%%%%%%%%%%%%%%%%%%%%%%%%%%%%%%%%%%%%%%%%%%%%
\section{Interaction of the branes}

Since the conformal invariance is preserved in the bulk action
(1), and broken on the boundary action (2), \cite{16}, the
conformal ghosts play role just in bulk and hence their
contribution in the boundary state should also be considered. For
calculating the interaction amplitude between two D-branes we
return to the previous boundary state (14) but restore the
integration over momenta. Those give the total boundary state
\begin{equation}
|B\rangle^{\rm total}=|B_{\rm gh}\rangle|B_{x}\rangle.
\end{equation} To find the
interaction amplitude between the D$p_{1}$ and D$p_{2}$-branes via
exchanging of closed string states, we need closed string propagator
which is given by time integral of the closed string Hamiltonian
\bea
&~& D=2\alpha'\int_{0}^{\infty}dt e^{-tH},
\nonumber\\
&~& H=\alpha'p^{\mu}p_{\mu}+2\sum^\infty_{n=1}(\alpha_{-n}.\alpha_{n}
+\widetilde{\alpha}_{-n}.\widetilde{\alpha}_{n})+(d-2)/6.
\nonumber
\eea

The convention for the indices which will be used in the amplitude is
as in the following. The set $\{\bar{i}\}$ shows the directions perpendicular
to both branes except $i_{0}$, $\{\bar{u}\}$ is for the directions
along both branes except $0$, $\{\alpha'_{1}\}$ is used for the
directions along the D$p_{1}$-brane and perpendicular to the
D$p_{2}$-brane, and $\{\alpha'_{2}\}$ indicates the directions along
the D$p_{2}$-brane and perpendicular to the D$p_{1}$-brane. Since
$\{\alpha_1\}$ and $\{\alpha_2\}$ are arbitrary, the position of the
branes are not fixed, that is, in our configurations the two branes
can be parallel or perpendicular to each other.
%%%%%%%%%%%%%%%%%%%%%%%%%%%%%%%%%%%%%%%%%%%%%%%%%%%%%%%%%%%%%%%%%
\subsection{The interaction amplitude}

The interaction amplitude is given by the overlap of the two boundary
state, corresponding to the branes, via the closed string propagator,
i.e. ${\cal{A}}=\langle B_{1}|D|B_{2}\rangle$. After a long calculation
we obtain
\begin{eqnarray}
{\cal {A}}=
&~&\frac{\alpha'V_{\overline{u}}}{4(2\pi)^{d_{\overline{i}}}}\frac{T
_{p_{1}}T_{p_{2}}}{|v_{1}-v_{2}|}\;[\det U_{1}\det
U_{2}]^{-1/2}\prod^{\infty}_{m=1}[\det R_{(m)1}\det R_{(m)2}]^{-1}
\nonumber\\
&~&\times{\int_{0}}^{\infty}dt \;
\bigg{\{}\prod^{\infty}_{m=1}\bigg{(}[\det(1-{\cal{S}}_{(m)1}
{\cal{S}}_{(m)2}^{T}e^{-4mt})]^{-1}(1-e^{-4mt})^{2}\bigg{)}
\nonumber\\
&~&\times
e^{(d-2)t/6}\bigg{(}\sqrt{\frac{\pi}{\alpha't}}\bigg{)}
^{d_{\bar{i}}}
\exp \bigg{(}-\frac{1}{4\alpha't}\sum_{\bar{i}}({y_{1}}^{\bar{i}}-
{y_{2}}^{\bar{i}})^{2}\bigg{)}
\frac{1}{\sqrt{\det Q \;\det G_{1}\;\det G_{2}}}
\nonumber\\
&~&\times \exp\bigg{(}-\frac{1}{4}\bigg{[}E^T Q^{-1}E
+\sum_{\alpha'_1}[({y_2}^{\alpha'_1})^2(G_1^{-1})^{\alpha'_1\alpha'_1}]
+\sum_{\alpha'_2}[({y_1}^{\alpha'_2})^2(G_2^{-1})^{\alpha'_2\alpha'_2}]
\bigg{]}\bigg{)}\bigg{\}}.
\end{eqnarray}
The matrices $Q$, $G_{1}$ and $G_{2}$ and the doublet $E$, are
defined through their elements as in the following
\begin{eqnarray}
\left\{
\begin{array}{rcl} &~& Q_{11}=
\frac{\alpha't}{(v_{2}-v_{1})^{2}}(1+{v_{1}}^{2})(1-{v_{2}}^{2})-
[(v_{1}^{2}+\frac{1}{2})^{2}(U^{00}_{1})^{-1}], \\
&~& Q_{22}=\frac{\alpha't}{(v_{2}-v_{1})^{2}}(1+{v_{2}}^{2})
(1-{v_{1}}^{2})-[(v_{2}^{2}+\frac{1}{2})^{2}(U^{00}_{2})^{-1}], \\
&~& Q_{12}=Q_{21}=\frac{\alpha't}{(v_{2}-v_{1})^{2}}(1+{v_{1}}^{2})
(1+{v_{2}}^{2})(1-v_{1}v_{2}),
\end{array}\right.
\end{eqnarray}
\begin{eqnarray}
\left\{ \begin{array}{rcl} &~&
E_{1}=\frac{i}{v_{2}-v_{1}}[{y_{2}}^{i_{0}}(1+{v_{1}}^{2})^{2}
-{y_{1}}^{i_{0}}(1+v_{1}v_{2})],
\\ &~& E_{2}=\frac{i}{v_{2}-v_{1}}[{y_{1}}^{i_{0}}
(1+{v_{2}}^{2})^{2} -{y_{2}}^{i_{0}}(1+v_{1}v_{2})],
\end{array}\right.
\end{eqnarray}
and the nonzero elements of the matrix $G_1$ are
\begin{eqnarray} \left\{ \begin{array}{rcl} &~&
G_{1\alpha'_1\alpha'_1}
=-\alpha't-\frac{1}{4}(U_{1}^{\alpha'_{1}\alpha'_{1}})^{-1}, \\
&~& G_{1\overline{u}\overline{u}}=-\frac{1}{2}\alpha't-
\frac{1}{4}(U_{1}^{\overline{u}\overline{u}})^{-1}.
\end{array}\right.
\end{eqnarray}
With the exchange $1\longleftrightarrow2$ we receive the nonzero
elements of $G_{2}$. Note that in (22) there is no sum on the repeated
indices $\alpha'_{1}$ and $\overline{u}$.

In the interaction amplitude (19), $V_{\overline{u}}$ is the
common worldvolume of the branes, and $d_{\bar i}$ is the
dimension of the directions which are perpendicular to both
branes. The infinite product in the second line of (19) shows the
effect of the oscillators and conformal ghosts, for analogue of it
see the Refs. \cite{9, 17}. The first exponential and its
pre-factor, which are emanated from the directions perpendicular
to both branes, indicate the damping of the amplitude due to the
distance of the branes. The momenta, which are in the Hamiltonian
and zero mode terms in boundary state, leads to the second
exponential and the pre-factor of it. The constant factors behind
the time integral, somehow show the strength of the interaction
which depend on the branes tensions, their velocities and the
tachyon fields. Note that the regularization of infinite product
in the first line can be done according to (12). Note that the
amplitude (19) can be interpreted as cylindrical partition
function for closed string, too.
%%%%%%%%%%%%%%%%%%%%%%%%%%%%%%%%%%%%%%%%%%%%%%%%%%%%%%%%%%%%%%%%%
\subsection{Long time behavior of the interaction amplitude}

One of the interesting features about the interaction amplitude is
to study its behavior after long enough times, i.e.
$\lim_{t\rightarrow\infty}{\cal {A}}$. In the ordinary cases (i.e.
in the absence of background tachyon) massless closed string states
dominate in this regime. Here the difference with the conventional
interaction amplitudes is the presence of the matrices $Q$, $G_{1}$,
$G_{2}$ and the doublet $E$, which are functions of time. Therefore,
in long distances of the branes in the $26$-dimensional spacetime,
the closed string tachyon and the massless closed string states
(i.e. graviton, dilaton and Kalb-Ramond) contribute to the
interaction amplitude as in the following
\begin{eqnarray}
{\cal {A}}_{0}
&~&=\lim_{t\rightarrow\infty}{\cal{A}}
\nonumber\\
&~&=\frac{i(-1)^{(p_{1}+p_{2})/2}
\;T_{p_{1}}T_{p_{2}}}{4(2\pi)^{d_{\bar i}}(1+{v_{1}}^{2})(1+{v_{2}}^{2})}\;\;
\frac{2^{d_{\overline{u}}+1/2}}{(\alpha')^{(p_{1}+p_{2})/2}}
\nonumber\\
&~&\times[\det U_{1}\det
U_{2}]^{-1/2}\prod^{\infty}_{m=1}[\det R_{(m)1}\det R_{(m)2}]^{-1}
\nonumber\\
&~&\times \int
dt\bigg{\{}\bigg{(}\sqrt{\frac{\pi}{\alpha't}}\bigg{)}
^{d_{\bar{i}}}
\exp \bigg{(}-\frac{1}{4\alpha't}\sum_{\bar{i}}({y_{1}}^{\bar{i}}-
{y_{2}}^{\bar{i}})^{2}\bigg{)}
\nonumber\\
&~& \times
\;\lim_{t\rightarrow\infty}\;\bigg{(}\frac{e^{4t}}{t^{1+(p_1+p_2)/2}}+
\frac{{\rm Tr}{({\cal{S}}_{(1)1}{\cal{S}}_{(1)2}^{T})-2}}
{t^{1+(p_1+p_2)/2}}\bigg{)}\bigg{\}},
\end{eqnarray}
where $d_{\bar u}$ is the dimension of the common worldvolume of
the branes. The limit of the exponential and its pre-factor in (23)
with respect to $t$ is not
important for us because they are related to the position of
the branes, while the states of closed string are independent of these
positions. The divergent part in the last line, (the first term),
corresponds to the tachyonic closed string state. The analogous of
this divergent term in the absence of the background tachyon field
lacks the decelerating coefficient $1/t^{1+(p_1+p_2)/2}$ and
is usually put away in the literature. It is a deficiency of
the bosonic string theory which will be recovered in superstring
theory. But the point is that here the time dependence in the
denominator slows down this divergence. The other term is related
to the contribution of the massless states which also differs from
the conventional case, due to the presence of the decelerating
factor which makes it to go to zero fast in the limit of long time.

There is a remarkable interpretation for this behavior. Taking
into account the open string tachyon as a background field, means
working with unstable D-branes. The consequence of this
instability is rolling of the tachyon as the system evolves and
after long time most of the energy which was localized in the
tachyon field transfers to bulk. This is the consequence of
decaying of the unstable D-branes into the bulk modes \cite{18}.
So, in this picture the long time interaction of the D-branes (due
to the massless closed string exchange) goes to zero. In other
words, after long enough time, there are no D-branes to consider
interaction for them. The exchange of the closed string tachyon
which is present as a divergent term also has been moderated in
this picture. Although this term tends to infinity anyway, its
rate of growth is related to the dimension of the branes.
Therefore, apart from the tachyonic term which goes to infinity,
we can say that the exchange of the massless closed string states
cause the D-branes to interact but their contribution decreases in
time due to the instability of the D-branes.

The damping of the interaction amplitude by passing the time
depends on the branes dimensions. An interesting exception is
D-instanton. When two D-instantons interact with each other
the factor $1/t^{1+(p_1+p_2)/2}$ reduces to $1$ and hence
the ordinary long time amplitude, associated with the massless
states, restores. In addition, is the usual divergent term
related to the tachyonic closed string state. So we can
say that the general interactive behavior of the D-instantons do
not change in the presence of background tachyon field.
%%%%%%%%%%%%%%%%%%%%%%%%%%%%%%%%%%%%%%%%%%%%%%%%%%%%%%%%
\section{Conclusions and Summary}

We obtained the boundary state of a closed string, emitted
(absorbed) from (by) a moving D$p$-brane in the presence of the
background tachyon field.

The relation between the boundary state and the
disk partition function is
discussed. The effect of the tachyon condensation on the partition
function was studied which shows a spectacular difference from
the conventional condensation.
Condensation of the tachyon matrix components along any
spatial worldvolume directions leads to a partition function
corresponding to a lower dimensional moving D-brane with an
effective tension which depends on the condensated components of
the tachyon field. However, condensation of $U_{00}$ eliminates
velocity and as well leads to an instantonic D-brane which is
fixed in time. After complete condensation of the tachyon field a
D-instanton is obtained.

The interaction amplitude between two D-branes with arbitrary
dimensions $p_1$ and $p_2$ was calculated. Our calculations are
valid for the systems that their branes are parallel or
perpendicular to each other. The interaction strength between the
branes depends on the branes dimensions, their tensions, their
relative configuration, the closed string mode numbers,
the tachyon matrices and the velocities of the branes.

As a special case, in the large distance interaction of the
branes, the contribution of the massless states goes to zero and
the divergence part related to the tachyonic state considerably
slows down. So the statement that the force associated with the
massless states is long range would be valid until there is no
tachyon background field in the system. This unconventional
behavior may be ascribed to the rolling of the tachyon field
towards its minimum potential. This leads to a closed string
vacuum without any D-brane at the end of the process and causes
the concept of interaction of the D-branes to faint. Interesting
point is that in the case of the D-instantons interaction
this descension of the long time amplitude jumps
to the usual case.
%%%%%%%%%%%%%%%%%%%%%%%%%%%%%%%%%%%%%%%%%%%%%%%%%%%%


\begin{thebibliography}{99}

\bibitem{1}
A. Sen, Int. J. Mod. Phys. A20 (2005) 5513-5656.
\bibitem{2}
A. Sen, Int. J. Mod. Phys. A14 (1999) 4061-4078.
\bibitem{3}
D. Kutasov, M. Marino and G. Moore, JHEP 0010 (2000) 045; T. Lee,
Phys. Lett. B 520 (2001) 385–390.
\bibitem{4}
K. Hashimoto, P. M. Ho and J. E. Wang, Mod. Phys. Lett. A20 (2005)
79-94.
\bibitem{5}
E. Cremmer and J. Scherk, Nucl. Phys. B50 (1972) 222.
\bibitem{6}
M.B. Green and P.Wai, Nucl. Phys.B431 (1994) 131;
M. Li, Nucl. Phys. B460 (1996) 351; C. Schmidhuber, Nucl. Phys.B467
(1996) 146; M.B. Green and M. Gutperle, Nucl. Phys. B476 (1996) 484;
P. Di Vecchia, M. Frau, I. Pesando, S. Sciuto, A. Lerda and R.
Russo, Nucl. Phys.B507 (1997) 259.
\bibitem{7}
J. Callan, C. Lovelace, C. R. Nappi and S. A. Yost, Nucl.
Phys. B308 (1988) 221; Nucl. Phys. B288 (1987) 525.
\bibitem{8}
Y. Zhang, Phys. Rev. D63 (2001) 106002; S. P. de Alvis, Phys.
Lett. B 505 (2001) 215–221; O. Andreev, Nucl. Phys. B598 (2001)
151-168.
\bibitem{9}
H. Arfaei and D. Kamani, Phys. Lett. B452 (1999)
54-60, hep-th/9909167; Nucl. Phys. B561 (1999) 57-76, hep-th/9911146.
\bibitem{10}
T. Lee, Phys. Rev. D64 (2001); G. Arutyunov, A. Pankiewicz
and B. Stefanski jr, JHEP 06 (2001) 049.
\bibitem{11}
E. T. Akhmedov, M. Laidlaw and G. W. Semenoff, JETP Lett. 77: 1-6,
2003, PismaZh. Eksp. Teor. Fiz. 77: 3-8, 2003; M. Laidlaw, G. W.
Semenoff, JHEP 0311:021, 2003; M. Laidlaw, ``{\it On a Modification of
the Boundary State Formalism in Off-shell String Theory}",
hep-th/0210270; T. Takayanagi, S. Terashima and T. Uesugi, JHEP
0103:019, 2001.
\bibitem{12}
D. Kutasov, M. Marino and G. Moore, ``{\it Remarks on Tachyon
Condensation in Superstring Field Theory}", hep-th/0010108; JHEP
0010 (2000) 045.
\bibitem{13}
P. Di Vecchia and A. Liccardo, NATO Adv. Study Inst. Ser. C. Math.
Phys. Sci. 556 (2000) 1-59.
\bibitem{14}
P. Kraus and F. Larsen, Phys. Rev. D63 (2001) 106004.
\bibitem{15}
E. S. Fradkin and A. A. Tseytlin, Phys. Lett. B163 (1985) 123.
\bibitem{16}
T. Uesugi, ``{\it Worldsheet Description of Tachyon
Condensation in Open String Theory}", hep-th/0302125.
\bibitem{17}
D. Kamani, Mod. Phys. Lett. A15 (2000) 1655-1664, hep-th/9910043.
\bibitem{18}
A. Sen, JHEP 0204 (2002) 048; F. Larsen, A. Naqvi, S. Terashima,
JHEP 0302 (2003) 039.

\end{thebibliography}
\end{document}